\documentclass[sigconf,nonacm]{acmart}

\usepackage{array}
\usepackage{tikz}
\usetikzlibrary{arrows.meta,positioning,fit,backgrounds}

\newcommand{\participantquote}[2]{%
  \begin{quote}
    \small
    \textbf{#1:} \textit{``#2''}
  \end{quote}
}
\setcopyright{none}

\acmConference[TAICHI 2026 Posters]{TAICHI 2026 Poster Track}{August 5--6, 2026}{Taipei, Taiwan}
\acmYear{2026}
\copyrightyear{2026}
\date{July 27, 2026}

\begin{document}

\title{What Makes an Initial Reaction Ready for Discussion?: Multi-Persona AI Support for Stance Reflection and Writing}

\author{Sky Shih-Kai Hong}
\email{sky.cs14@nycu.edu.tw}
\affiliation{%
  \institution{National Yang Ming Chiao Tung University}
  \city{Hsinchu}
  \country{Taiwan}
}

\author{Mu-Tien Kuo}
\email{mtkuo.cs13@nycu.edu.tw}
\affiliation{%
  \institution{National Yang Ming Chiao Tung University}
  \city{Hsinchu}
  \country{Taiwan}
}

\author{Wei-Ji Chen}
\email{umineko.cs13@nycu.edu.tw}
\affiliation{%
  \institution{National Yang Ming Chiao Tung University}
  \city{Hsinchu}
  \country{Taiwan}
}

\author{Dennis Wang}
\email{denniswang@cs.nycu.edu.tw}
\affiliation{%
  \institution{National Yang Ming Chiao Tung University}
  \city{Hsinchu}
  \country{Taiwan}
}

\renewcommand{\shortauthors}{Hong et al.}

\begin{abstract}
An initial reaction to a social or community issue can feel meaningful before it is ready to become a message: people still need to clarify the claim, anticipate audience risks, and decide how much reasoning should become visible to others. We present StanceLab, a prototype for preparing a stance before entering a discussion. The prototype compares a three-persona mode, where an Interviewer, Mentor, and Opponent respond in parallel to help users diagnose and revise a stance, with a standalone LLM mode. In a formative within-subject pilot with six participants and 12 task sessions, every session produced a short final message in the notepad. The pilot revealed two design requirements: persona roles should diagnose useful blind spots or objections, and parallel responses need coordination support. We propose a future diagnosis-and-writing workflow that turns persona-based reflection into selective, audience-aware final messages.
\end{abstract}


\begin{teaserfigure}
  \centering
  \begin{tikzpicture}[
    font=\sffamily,
    node distance=0.1cm and 0.46cm,
    box/.style={
      draw=black!55,
      rounded corners=2pt,
      line width=0.55pt,
      minimum height=1.15cm,
      align=center,
      fill=black!3
    },
    role/.style={
      draw=black!45,
      rounded corners=2pt,
      line width=0.45pt,
      minimum height=0.56cm,
      text width=3.55cm,
      align=left,
      font=\scriptsize\sffamily,
      fill=blue!4
    },
    arrow/.style={-{Latex[length=2mm]}, line width=0.55pt, draw=black!65},
    looparrow/.style={-{Latex[length=1.8mm]}, line width=0.45pt, draw=blue!55},
    note/.style={font=\scriptsize\sffamily, align=center, text=black!70}
  ]
    \node[box, text width=3.05cm, fill=red!5] (reaction) {
      \textbf{Initial reaction}\\
      \scriptsize Meaningful reaction needing discussion prep
    };
    \node[role, right=0.72cm of reaction] (mentor) {\textbf{Mentor:} organize reasons and structure};
    \node[role, above=0.07cm of mentor] (interviewer) {\textbf{Interviewer:} clarify meaning and context};
    \node[note, above=0.06cm of interviewer] (diagnosisTitle) {\textbf{Persona diagnosis}};
    \node[role, below=0.07cm of mentor] (opponent) {\textbf{Opponent:} surface objections and risks};
    \node[box, text width=3.65cm, right=0.62cm of mentor, fill=green!5] (draft) {
      \textbf{Stance workspace}\\
      \scriptsize Select useful points, revise in personal voice, choose disclosure level
    };
    \node[box, text width=3.15cm, right=of draft, fill=orange!7] (ready) {
      \textbf{Shareable}\\
      \textbf{final message}\\
      \scriptsize Clear to others before sharing
    };
    \begin{scope}[on background layer]
      \node[
        draw=blue!35,
        dashed,
        rounded corners=4pt,
        line width=0.45pt,
        inner xsep=0.16cm,
        inner ysep=0.22cm,
        fit=(diagnosisTitle)(interviewer)(opponent)(draft)
      ] (loopFrame) {};
    \end{scope}
    \node[note, fill=white, inner xsep=2pt, inner ysep=0pt] at (loopFrame.north) {\textbf{Iterative reflection-writing loop}};
    \draw[arrow] (reaction.east) -- (mentor.west);
    \draw[looparrow] (mentor.east) to[bend left=28] (draft.west);
    \draw[looparrow] (draft.west) to[bend left=28] (mentor.east);
    \draw[arrow] (draft) -- (ready);
  \end{tikzpicture}
  \caption{Our pilot motivates an iterative StanceLab workflow that connects persona-based diagnosis with stance writing: persona roles diagnose what an initial reaction still needs (missing context, structure, objections), and a writing workspace turns the selected points into a documented, shareable stance.}
  \Description{A teaser diagram showing the workflow motivated by the pilot. It starts with an initial reaction, enters an iterative reflection-writing loop that includes persona diagnosis by Interviewer, Mentor, and Opponent roles plus a stance workspace, and ends with a shareable final message.}
  \label{fig:workflow}
\end{teaserfigure}
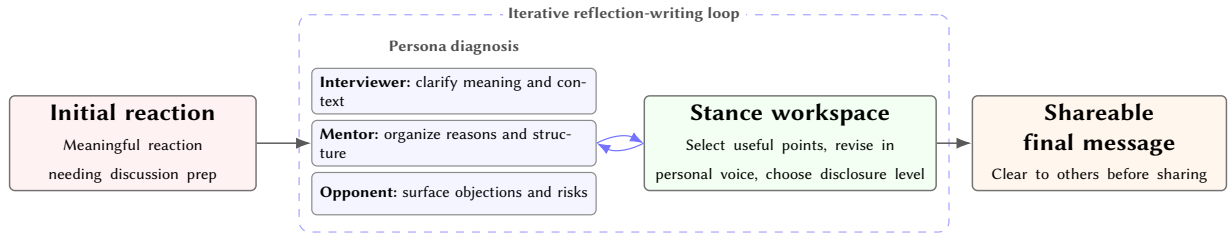

\maketitle

\begin{figure*}[!t]
  \centering
  \begin{minipage}[t]{0.275\textwidth}
    \centering
    \includegraphics[height=0.308\textheight,trim=250 0 250 0,clip]{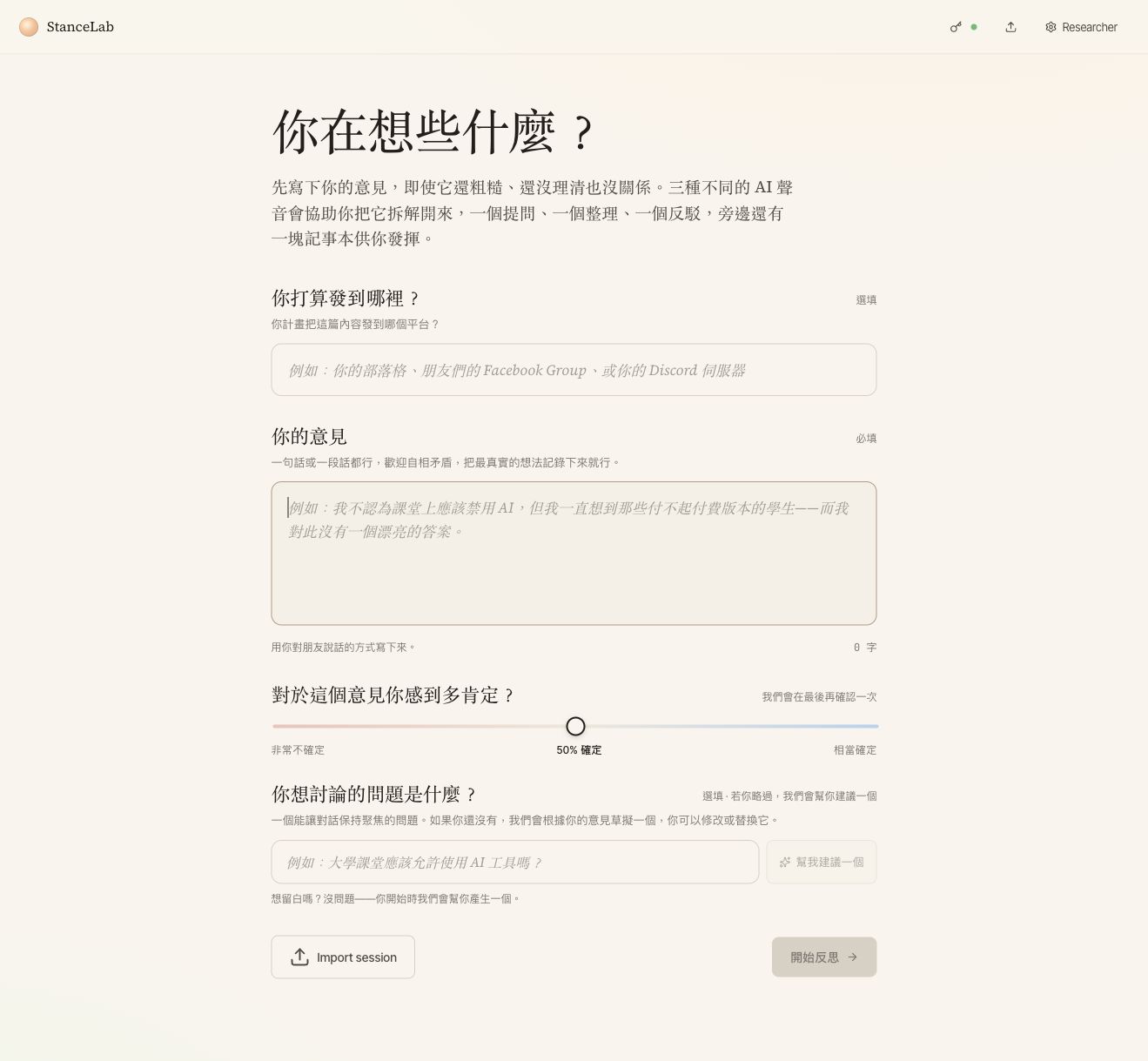}
    \caption*{\small (a) Setup: audience, issue, stance, and confidence.}
  \end{minipage}
  \hfill
  \begin{minipage}[t]{0.44\textwidth}
    \centering
    \includegraphics[height=0.308\textheight]{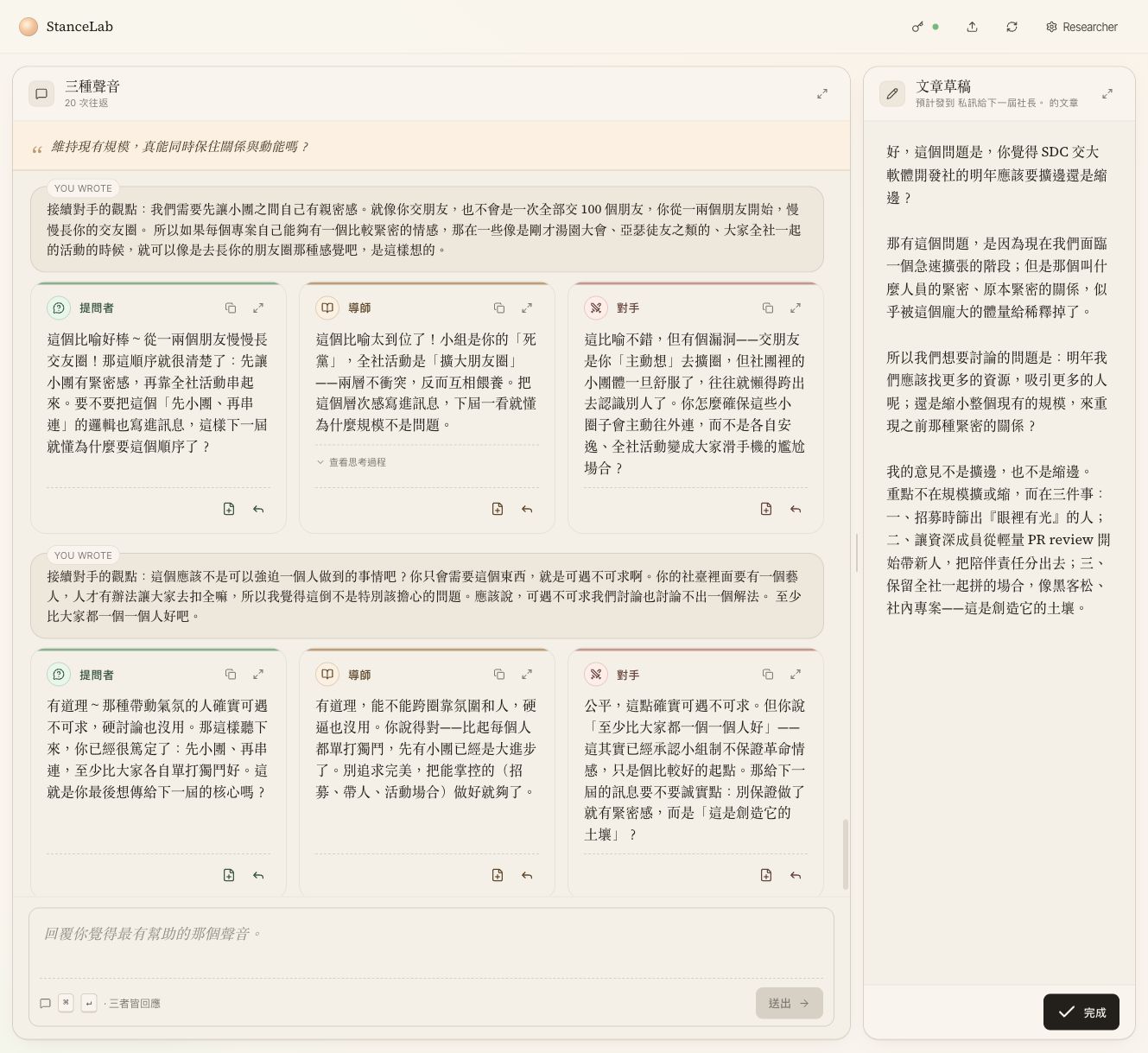}
    \caption*{\small (b) Three-persona condition: roles respond in parallel.}
  \end{minipage}
  \hfill
  \begin{minipage}[t]{0.275\textwidth}
    \centering
    \includegraphics[height=0.308\textheight,trim=260 0 260 0,clip]{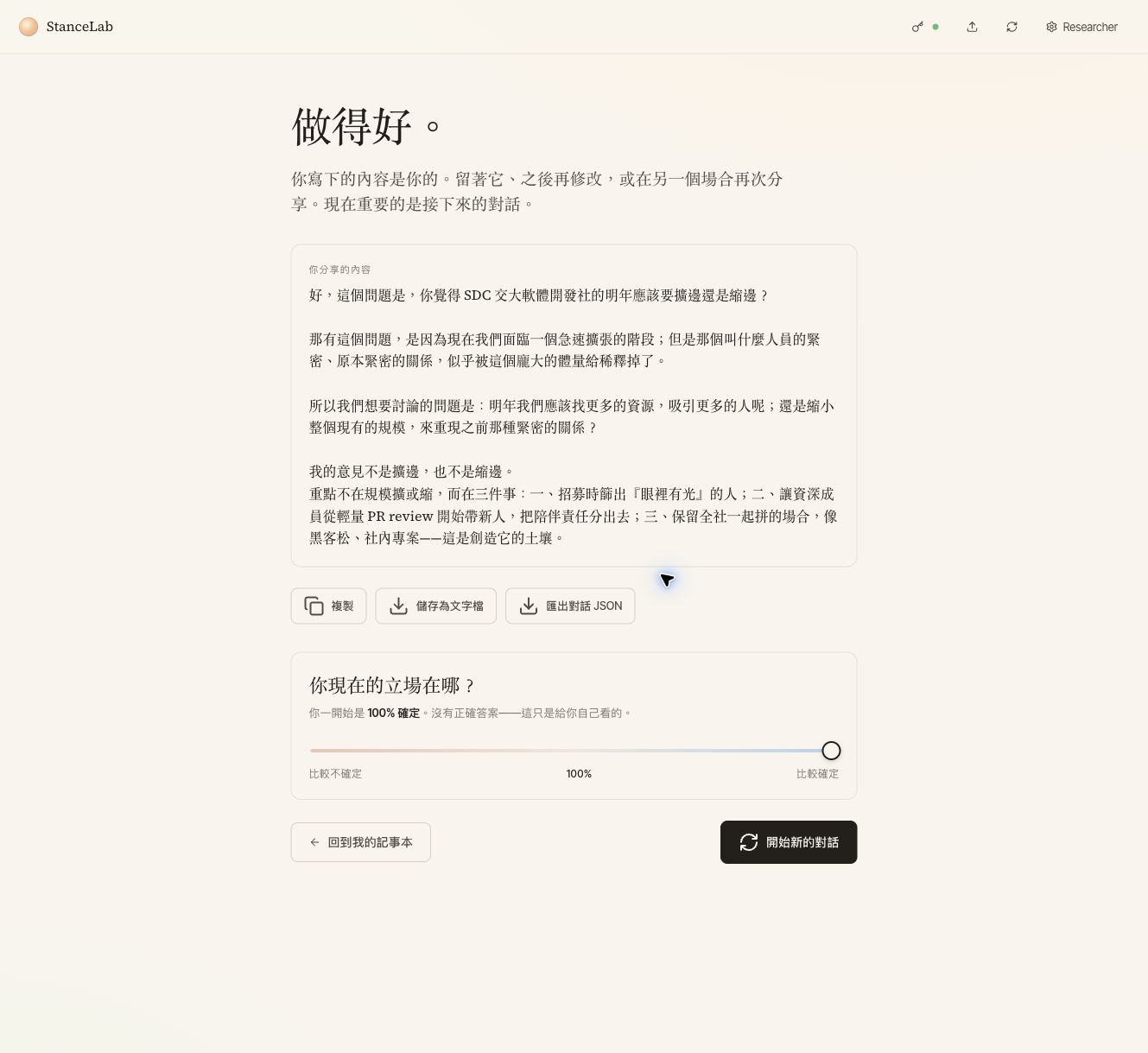}
    \caption*{\small (d) Final stage: review summary and write a message.}
  \end{minipage}
  \caption{Current StanceLab interface across setup, three-persona discussion, and final review stages.}
  \Description{Three screenshots of StanceLab. The first shows the setup page for issue, opinion, and confidence. The second shows the three-persona discussion interface with three persona roles. The third shows the final stage with a discussion summary and final-message area.}
  \label{fig:interface}
\end{figure*}

\section{Motivation and Objective}
Online discussions about social and community issues often move quickly from emotion to assertion. A meaningful initial reaction still requires preparation before discussion. Before sharing in social or community discussion spaces, people may need to clarify what they believe, identify weak assumptions, anticipate how others may challenge the claim, and decide how much reasoning should become visible to others. We define stance preparation as the work of turning an initial reaction into a claim that can be discussed, revised, or selectively shared. Writing is central to this work: documenting ideas and stances---in drafts, notes, or unsent replies---is already a common step before people join online discussions, and the written stance is what ultimately gets shared. This preparation involves at least three recurring moves: clarification, organization, and challenge. It also involves writing decisions: how to preserve personal voice, how to package reasoning for an audience, and what should remain private. This poster paper asks how conversational AI can support this pre-deliberation work while preserving the user's ownership of the final post. The intended contribution is practical and formative: the pilot reveals design requirements for AI-assisted stance preparation. Persona diversity is useful only when it helps users notice something actionable, and reflection support needs to be connected to draft revision and audience choice.

\section{Related Work}
This work builds on reflection support systems, opinion exploration, and conversational-agent role design. Prior systems show that conversational prompts can scaffold reflection in everyday and workplace contexts~\cite{kocielnik2018workplace,sakel2024socialjournal}. Work on opinion exploration and deliberation shows the value of structured exposure to multiple perspectives~\cite{kim2021starrythoughts,khadpe2022empathosphere,zhang2023deliberating}. Conversational-agent research further suggests that an agent's role and communication style can change the user experience~\cite{poivet2023conversational}. We combine these ideas in a narrower setting: preparing a personal stance before a user decides whether and how to join a discussion. Here, diagnosis means helping users identify under-specified claims, missing context, audience risks, and useful objections before writing.

\section{Prototype}

We built StanceLab\footnote{Source code: \url{https://github.com/skyhong2002/StanceLab}.} as a SvelteKit front-end prototype using Vite, TypeScript, Tailwind CSS, and OpenRouter API integration. The prototype follows a preparation workflow: an initial reaction enters a reflection-writing loop, persona diagnosis surfaces issues, and a workspace turns selected points into a shareable message. Here, diagnosis and writing play distinct roles: the personas diagnose what the stance still needs---missing context, weak structure, unaddressed objections---while the writing workspace is where the user turns selected diagnoses into a documented stance. We use three distinct personas, rather than one general assistant, because each persona instantiates one of the three preparatory moves introduced above, keeping the moves separate and simultaneously visible. The Interviewer prompt asks one focused, non-leading question to surface missing context; the Mentor prompt offers one organizing insight or concrete suggestion; and the Opponent prompt gives one respectful but sharp challenge or edge case.

\begin{figure}[!b]
  \centering
  \includegraphics[width=\linewidth,height=0.24\textheight,keepaspectratio]{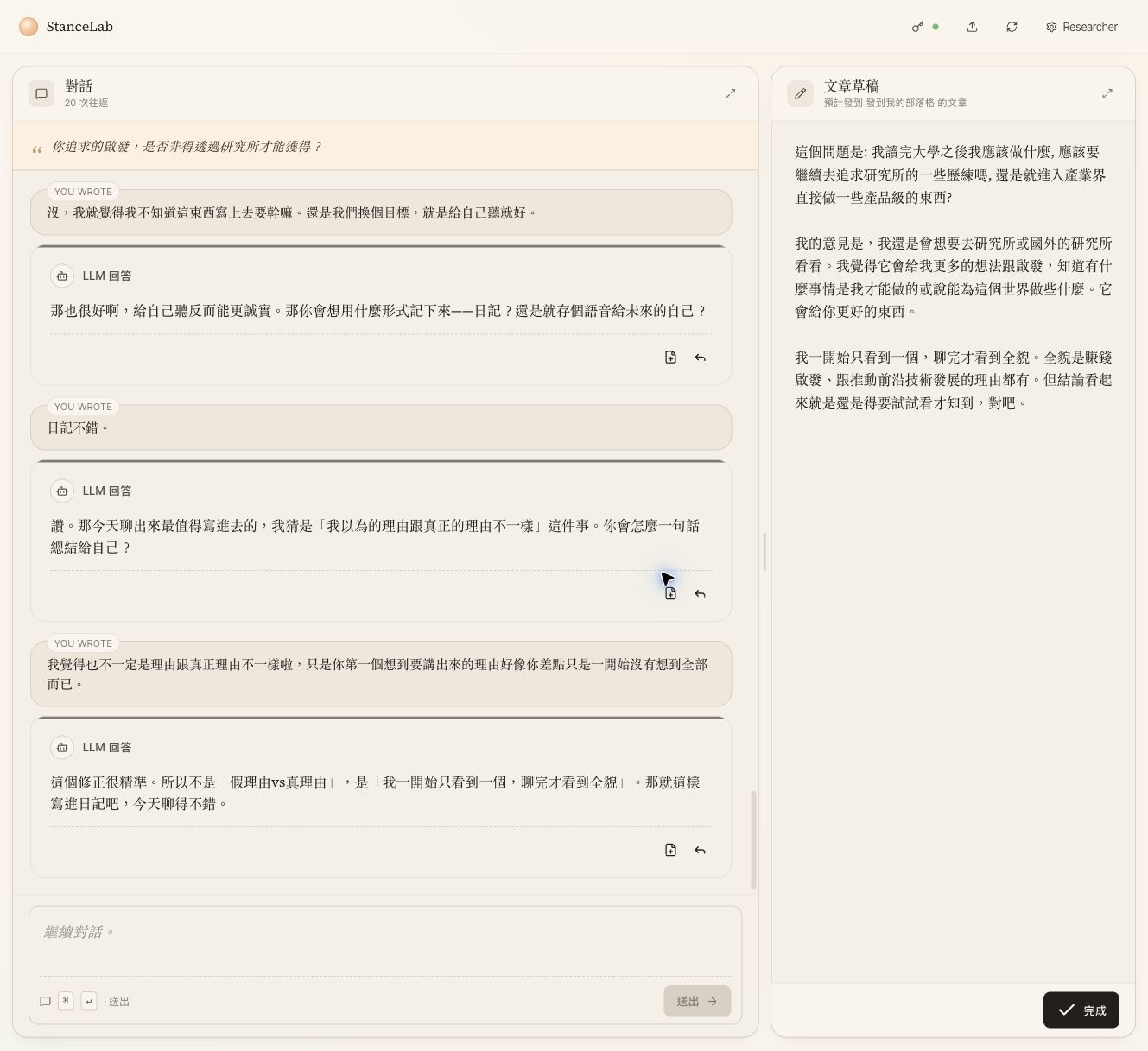}
  \caption{(c) Standalone LLM control condition: one conversational thread supports reflection under the same task framing.}
  \Description{A screenshot of the standalone LLM condition in StanceLab, showing a single conversational thread and a draft area.}
  \label{fig:standalone}
\end{figure}

The study version includes two modes. In the three-persona mode, each user message is sent to three role prompts in parallel; each prompt receives the same user context and prior responses from the other roles, stays in role, and focuses on the user. Users can inspect the persona responses, continue the conversation, reply to a persona role, and move material into a draft area. In the standalone mode, users interact with a single LLM under the same broad task framing, as shown separately in Figure~\ref{fig:standalone}.

Both modes begin by asking for the intended sharing context, initial opinion, confidence level, and issue to explore. Figure~\ref{fig:workflow} shows how the three-persona mode turns this shared context into persona diagnosis: the Interviewer, Mentor, and Opponent respond in parallel, and the user selects useful points for revision in the stance workspace. This setup keeps the conversation anchored to a possible audience and written output. The visible draft area positions the system as a tool for moving from reflection to a short written stance.

\section{Method}

We conducted a formative within-subject pilot study with six participants recruited from the student developer community. Each participant completed two stance-reflection tasks, using both the three-persona mode and the standalone mode. One task used a shared community prompt about whether a student developer community should keep expanding or reduce its scale to rebuild interpersonal ties. The other task asked participants to choose a personally relevant issue they would naturally want AI help thinking through; participants brought issues such as study or career decisions, organizational messages, and AI development. This mix let us observe both community-facing and more personal reflection, but it also means our mode comparison is descriptive: topic sensitivity and task framing varied across sessions. Mode order was counterbalanced across participants.

Each session followed a moderated protocol. Participants stated an initial stance, rated confidence on a 0--100 slider, conversed with the system, and produced a short final message in the notepad. Data sources included session JSON logs, final notepad outputs, exported post files for a subset of sessions, audio recordings, interview transcripts, AI-generated interview summaries, and observation notes. For this poster, we computed descriptive summaries from the JSON logs and notepad fields: turn counts, final-message completion, and pre/post confidence changes. We reviewed transcripts, summaries, notes, and drafts for recurring critical incidents, including blind-spot diagnoses, coordination load, role successes or failures, writing moves, and sharing decisions. We grouped these incidents into design findings and report participant counts for each pattern.

\begin{table}[t]
  \caption{Pilot outcome summary. Study overview: six participants completed 12 sessions, producing 146 logged user turns, 12 final notepad messages, and 9 sessions with post-confidence ratings.}
  \label{tab:dataset}
  \small
  \begin{tabular}{p{0.38\linewidth}p{0.46\linewidth}}
    \toprule
    Outcome & Descriptive result \\
    \midrule
    Final-message completion & 12 / 12 sessions ended with a final notepad message. \\
    Mean confidence change & +16.8 points among sessions with post-confidence. \\
    Confidence direction & 7 sessions rose, 2 held, and 0 fell among recorded post-session ratings. \\
    \bottomrule
  \end{tabular}
\end{table}

\section{Current Results}

All 12 sessions ended with a short final message in the notepad, which suggests that the task framing successfully connected conversation to writing. The 146 user turns show engagement beyond one-shot prompting. The standalone mode averaged more turns than the three-persona mode, so engagement patterns require cautious descriptive interpretation. Post-session confidence means the participant's 0--100 self-rating after completing the conversation and final message. It rose in most recorded sessions. Because it was missing in three sessions and modes were paired with different topics, we treat this as descriptive evidence that participants often left with a more articulated stance.

\subsection{Persona Roles Helped When They Diagnosed a Blind Spot}

The strongest value of the three-persona mode came from relevant questions, objections, and redefinitions. At least three participants described value in moments where a persona interrupted their current frame with a usable diagnostic move. P01 described the value of challenge directly:

\participantquote{P01}{Isn't this the point of debate? If you just follow your own thinking, you can actually complete that process alone; you do not need another person to discuss it with you. Because two people are, by default, here to discuss this.}

P05 made a similar point in the shared community task. A helpful response asked whether the real problem was community headcount or a more specific cause:

\participantquote{P05}{Are you talking about the first one? The three-person one? It produced more words... no, I do not think it was because of the word count. It was the question it raised. It really asked a question, and it hit that point.}

Together with P06's request for personas to diagnose problems before an integrated revision step, these cases suggest that multi-persona AI should be evaluated by whether at least one response produces an actionable diagnosis: a missing context question, a weak assumption, an objection, or a useful reframe.

This result reframes the value of persona diversity. A persona response became useful when it changed what the participant could do next. In several sessions, the useful moment was a narrow diagnostic hit: a question that exposed an undefined term, an objection that forced the participant to name evidence standards, or a mentor-like summary that converted scattered thoughts into a draftable structure.

\subsection{Parallel Responses Also Created Coordination Load}

The same design also created friction. Three participants described some form of coordination load when multiple persona responses remained separate. P06 said that the three-persona mode felt like:

\participantquote{P06}{It felt a bit like having to talk to three people at once... I normally cannot suddenly handle that many things at once. I tend to reply to one at a time, and that can already take 70--80\% of my attention.}

P03 described the interaction as one question-answer session becoming three separate question-answer sessions:

\participantquote{P03}{The difference between the two modes was just that one session became two sessions, three sessions, like that. In essence, there was not much difference.}

P02 further showed that multiplicity depends on topic and audience. For personal topics, three persona responses could feel less like help and more like judgment:

\participantquote{P02}{For the first mode, it felt more like I had posted something on Facebook and was being publicly judged. Then people with three different stances came to talk to me.}

Participants often valued multiple perspectives and wanted the system to integrate the persona responses before asking the user to respond. This also changes how we interpret reply logs: replying to a persona role may indicate usefulness, confusion, or the need to correct a misunderstanding.

\subsection{Persona Role Value Was Uneven}

At least three participants explicitly contrasted the individual roles. The Interviewer was useful when it asked a specific clarification question; P03 felt it could become vague agreement or overlap with the Mentor. The Mentor was more stable because it organized context and gave improvement directions; P06 described it as first understanding the background and then offering a challenge. The Opponent was high-risk and high-reward: it helped when it challenged a relevant weak point; P03 said an off-topic challenge would ``only waste tokens.''

\begin{table}[t]
  \caption{Observed persona-role patterns.}
  \label{tab:roles}
  \small
  \begin{tabular}{p{0.22\linewidth}p{0.32\linewidth}p{0.32\linewidth}}
    \toprule
    Persona role & Worked when & Broke when \\
    \midrule
    Interviewer & Asked a specific question that clarified context or assumptions. & Produced vague agreement or overlapped with the Mentor. \\
    Mentor & Organized the stance and named a concrete next writing move. & Became reassurance while leaving the argument unchanged. \\
    Opponent & Challenged a relevant weak point or alternative interpretation. & Used weak analogies or argued against a point the user had already moved past. \\
    \bottomrule
  \end{tabular}
\end{table}

Table~\ref{tab:roles} summarizes why the next design should evaluate persona roles by their function in the user's workflow. A persona label needs to produce a distinct, actionable move. The future system should be able to explain why a persona role's output matters: what assumption it found, what risk it predicts, or what revision it enables.

\subsection{Reflection Needed a Path Into Writing}

Three participants explicitly pointed toward a writing workflow beyond a pure chat tool. P04 copied useful AI points into the draft and then rewrote tone, punctuation, and transitions:

\participantquote{P04}{Most of the time, I had discussed with it up to a point, and it had produced a viewpoint I thought I could adopt. I pasted it over first. But actually, I later found that I could directly ask it to rewrite for me.}

P06 wanted the three personas to identify problems first, then have one integrated LLM revise the text with the full prior context:

\participantquote{P06}{After I first type something, I would feed it to the three personas first, let them each point out my problems. Then I would use an integrated LLM to improve that pile of problems one by one through interaction.}

P02 added that public-facing writing should preserve personal voice and limit full AI generation:

\participantquote{P02}{If it is a post that will go out, I basically would not let AI participate too much in the content I write. Mainly, it would only help me fix typos... I would try to keep more of my own thoughts, my thoughts at that moment.}

These observations point toward a selective workflow: keep this point, rewrite it in my voice, and package it for this audience should be separate actions.

\subsection{Sharing Was Selective}

The final messages often served purposes beyond immediate posting. Three participants distinguished between private preparation and public expression. P01 described one output as something kept for the self:

\participantquote{P01}{I do not know. I normally do not post, so I probably would not send it to anyone. I would just treat it as my own.}

P02 made a similar distinction between public conclusions and private work:

\participantquote{P02}{It is like a math solution: I give you the formula and the answer, and the mistaken calculation process in the middle... especially for something very public, that is not something people particularly care about, or something I want others to see.}

P05 showed the other side of the finding. For one community-related task, the participant said the output could be sent to a leader because the system had added concrete implementation details:

\participantquote{P05}{With it, I would feel more that I should send it, because it has more information and can lead to a more meaningful discussion... besides expressing an opinion, there are also some executable actions.}

This matters because pre-discussion reflection can contain uncertainty, rejected ideas, sensitive context, and half-formed reasoning. StanceLab should therefore help users choose a disclosure level for each final message.

\section{Limitations and Future Research Plan}

This work uses a small formative pilot for design exploration. Participants came from a nearby student developer community, so their AI-tool familiarity and discussion habits may differ from broader publics. The within-subject comparison also used different topics with varied sensitivity and audience, and post-session confidence was recorded in 9 of 12 sessions. We therefore use the findings as design evidence for the next prototype and leave general deliberation-quality claims for future controlled studies.

Future work should measure whether participants share final messages, how recipients respond, and how prepared stances shape downstream discussion while preserving the privacy boundary that makes preparation useful. The next version of StanceLab will treat personas as a diagnostic layer with an integrated writing path: identify the strongest claim, missing context, likely audience misunderstanding, and useful objections; synthesize these into a compact diagnosis card; and support a workspace for saving AI points, converting critiques into revision checklists, preserving personal voice, and choosing disclosure levels such as private note, friend message, club post, or public argument.

\section{Conclusion}

StanceLab shows that multi-persona AI can support stance preparation when its persona roles surface a useful blind spot, question, or challenge. The pilot also shows that parallel persona responses create work when they remain separate from drafting. The most promising direction is therefore a diagnosis-and-writing workflow: use persona diversity to identify what needs attention, then help the user selectively turn those insights into a personally controlled, audience-appropriate stance.

\begin{acks}
This project was developed as coursework for CSIC30196: Social Computing and Social Media Application Design (114-2 / Spring 2026) at National Yang Ming Chiao Tung University. We thank the pilot participants for their feedback and participation.
\end{acks}

\bibliographystyle{ACM-Reference-Format}
\bibliography{references}

\end{document}